\documentclass[journal]{IEEEtran}

\usepackage{graphicx}
\usepackage{amsfonts}
\usepackage{amsmath}
\usepackage{amssymb}
\usepackage{color}
\usepackage{cite}
\usepackage{graphicx,color,overpic}
\usepackage{psfrag}
\usepackage{amssymb}
\usepackage{times}
\usepackage{latexsym}
\usepackage{bm}
\usepackage{cases}
\usepackage{array}
\usepackage{fancyhdr}
\usepackage{setspace}
\usepackage{subfigure}
\usepackage{url}
\usepackage{multirow}
\usepackage{epstopdf}

\usepackage{epsfig}
\usepackage{fancybox}
\usepackage{textcomp}

\DeclareMathOperator*{\argminA}{arg\,min} 

\newcommand{\SIM}[0]{\mathrm{sim}}
\newcommand{\x}[0]{\mathbf{x}}
\newcommand{\z}[0]{\mathbf{z}}
\newcommand{\F}[0]{\mathbf{F}}
\newcommand{\f}[0]{\mathbf{f}}

\newcommand{\s}[0]{\mathbf{s}}

\newcommand{\binning}[0]{\mathbf{b}}

\title{\LARGE \bf
Collaborative SLAM based on Wifi Fingerprint Similarity and Motion Information
}

\author{Ran Liu, Sumudu Hasala Marakkalage, Madhushanka Padmal, Thiruketheeswaran Shaganan, \\
Chau Yuen, Yong Liang Guan, and U-Xuan Tan
\thanks{This work is partly supported by the National Science Foundation of China (No. 61601381, 61750110529, and 61701421) and the Sichuan Science and Technology Program (No. 2019YFH0161 and 2019JDTD0019).} 
\thanks{R. Liu, S. H. Marakkalage, C. Yuen, and U-X. Tan are with the Engineering Product Development Pillar, Singapore University of Technology and Design, 8 Somapah Rd, Singapore, 487372. 
R. Liu is also with the School of Information Engineering, Southwest University of Science and Technology, 59 Qinglong Road, Mianyang, China, 621010.
{\{\tt\small ran\_liu, yuenchau, uxuan\_tan\}@sutd.edu.sg}.
}
\thanks{M. Padmal and T. Shaganan are with the Department of Electronic and Telecommunication Engineering, University of Moratuwa, Sri Lanka, 10400.
}
\thanks{Y. L. Guan is with the School of Electrical and Electronic Engineering, Nanyang Technological University, 50 Nanyang Avenue, Singapore, 639798
{\tt\small eylguan@ntu.edu.sg}. 
}
}

\begin{document}

\maketitle
\thispagestyle{empty}
\pagestyle{empty}

\begin{abstract}
Simultaneous localization and mapping (SLAM) 
has been extensively researched in past years particularly with regard to range-based or visual-based sensors.
Instead of deploying dedicated devices that use visual features, 
it is more pragmatic to exploit the radio features to achieve this task, 
due to their ubiquitous nature and the widespread deployment of Wi-Fi wireless network.
This paper presents a novel approach for collaborative simultaneous localization and radio fingerprint mapping (C-SLAM-RF) in large unknown indoor environments. 
The proposed system uses 
received signal strengths (RSS) from Wi-Fi access points (AP) in the existing infrastructure 
and pedestrian dead reckoning (PDR) from a smart phone, 
without a prior knowledge about map or distribution of AP in the environment. 
We claim a loop closure based on the similarity of the two radio fingerprints.
To further improve the performance, 
we incorporate the turning motion 
and assign a small uncertainty value to a loop closure if a matched turning is identified. 
The experiment was done in an area of 130 meters by 70 meters and the results show that our proposed system 
is capable of estimating the tracks of four users with an accuracy of 0.6 meters with Tango-based PDR
and 4.76 meters with a step counter-based PDR. 

\end{abstract}

\section{Introduction}
\label{Introduction}
With growing applications of the Internet of Things (IoT), 
recent research shows an increasing interest in indoor positioning due 
to the rapid demand of location-based services, 
such as indoor guidance and asset tracking \cite{Yassin_ieee_tutorials_2016, he2016wi, wifi_fusion, localization_framework}. 
To perform indoor positioning, 
the knowledge of the existing infrastructure must be provided in advance 
(for example a map of the environment or locations of the beacons). 
In the scenario of emergency response in disaster areas
or large scale environments, 
such kind of knowledge is not available or difficult to obtain beforehand, 
which makes the indoor positioning challenging. 
Therefore, 
recent researchers are focusing on developing efficient methods and technologies 
to simultaneously localize mobile devices (robots and smartphones) and generate a map of the environment \cite{montemerlo2002fastslam,ferris2007wifi,burgard2009comparison,survey_av_localization}. 
The underlying problem is well known by the term, Simultaneous Localization and Mapping (SLAM). 
Extensive researches have been done with regard to visual-based \cite{Taketomi2017} or range-based sensors \cite{slam_trends_2016}. 

Loop closure detection is elementary to any SLAM system. 
It denotes a situation that the mobile device has entered a previously-visited location, 
which permits to correct the accumulated odometry error. 
In order to perform loop closure detection in SLAM, 
dedicated devices (i.e., laser range finders or cameras) are required to measure the similarity of observations by 
scan matching \cite{Lu_millos_1997} or feature matching \cite{Taketomi2017}, which are usually computationally expensive. 
However, 
growing popularity of Wi-Fi wireless networks provide a new opportunity to detect loop closure and perform SLAM in a different way. 



Most existing buildings with Wi-Fi network deployed can be exploited for localization and mapping with low hardware requirements and computational cost 
(for example with the ubiquitous IoT devices like normal smartphones) due to their ubiquitous nature of in-built sensing capabilities \cite{Yassin_ieee_tutorials_2016, ran_ieee_sensors2017}. 
The current signal-strength-based SLAM requires an analytical model to feature the radio signal distribution over distance \cite{ferris2007wifi, Huang_wifi_slam_11}. 
However, it is not practical to build such a model due to many multiple path issues in uncontrolled environments. 
On the other hand, radio fingerprinting \cite{he2016wi, pervasive_integration_imu_fp, Robust_neighborhood_graphing}, 
represents a location with a collection of radio signals from Wi-Fi access points, 
which is considered to be more robust against the signal distortions. 
Therefore, we adopt this technique to simultaneously determine the location of a user and create a radio map of the environment.

In addition to the Wi-Fi network, a typical indoor environment consists of many landmarks, such as turnings, elevators, rooms, and doors, 
which can be also considered as features for the positioning of a device.
These landmarks can be recognized through inertial sensors, 
which are available in most commercial off-the-shelf smartphones \cite{turn_detection_trans, kaiser_elevator_detection_2016}.
In contrast to large 
location uncertainty of radio fingerprints due to the distortion of signals, 
such kind of landmarks can better confine the location of a device and enhance the positioning accuracy of fingerprinting-based approaches \cite{Robust_neighborhood_graphing, wang_motion_slam_iros}.

In opposite to the feature map or occupancy map built by laser range finders or visual cameras, 
our goal is to build a map (in particular a radio map) with radio fingerprint as feature, and use that for the positioning.
To ensure a good positioning accuracy in large scale environment, 
a fine-grained radio map is required \cite{Yassin_ieee_tutorials_2016, he2016wi} and it will be time consuming to create such a map with a single user.
Therefore, a low cost method (e.g., acquire fingerprints via crowdsensing by multiple users) to create the radio map is a necessity. 


This paper presents a system that fuses the pedestrian dead reckoning from a smartphone, 
and received signal strength (RSS) measurements from surrounding Wi-Fi access points (AP), 
to estimate the trajectory of multiple users and map the radio signals in unknown environment via a collaborative fashion, 
using graph SLAM technique.
To further improve the accuracy, 
we incorporate the turning features and 
reduce the uncertainty of loops inferred based on radio fingerprints similarity.
The proposed approach requires neither the map of the environment nor the locations of the access points. 
We tested the system under two different dead reckoning systems, 
one is based on Tango that has a high motion tracking accuracy through vision-based odometry, 
and the second one is based on step counter using on-board inertial sensors that has a poor motion tracking accuracy. 
By leveraging on in-built sensing capabilities from smart phones and crowdsensing nature, 
our system can generate a radio fingerprint map in a large indoor environment at low cost as compared with traditional site surveying methods.

We summarize the contributions of this paper as:
\begin{itemize}
\item We present a solution that incorporates Wi-Fi fingerprint and dead reckoning information for crowdsensing SLAM in unknown indoor environments;
\item We propose an algorithm that automatically learns a model to characterize the uncertainty of a loop based on the degree of similarity using the short term odometry measurement;
\item We integrate the turning features to further reduce the uncertainty of radio fingerprint-based loop closures and improve the overall accuracy;
\item We throughly evaluate our approach in one building at our campus with an area of approx. 9000 square meters with two different pedestrian dead reckoning systems.
\end{itemize}

We organize the rest of this paper as follows.
The related work is discussed in Section \ref{related_work}.
Section \ref{system_overview} formulates the problem and explains the detail of the proposed system. 
Section \ref{experimental_evaluations} presents the experimental results. 
Conclusions with possible directions of future work are made in Section \ref{conclusions}.

\section{Related Work}
\label{related_work}
Over the past decades, indoor positioning shows a growing popularity 
due to the increasing demand of location-aware applications \cite{Yassin_ieee_tutorials_2016, he2016wi}.
A large number of researches have been performed 
regarding indoor positioning given a reference of the infrastructure (i.e., map of the environment or distributions of beacons).
Obtaining and maintaining such kind of information is challenging, 
particularly in large scale environments \cite{wifi_positioning_challenge} or emergency response for example search and rescue in disaster scenes \cite{Liu_Relative_Globecom, Liu_relative_positioning_icra}.
A solution to this problem is SLAM (Simultaneous Localization and Mapping), 
which has been investigated extensively in robotics community.
In this section, we present a summary of the related work in SLAM, 
using different kinds of techniques. 
Throughout the years, many techniques and algorithms
have been proposed, mainly including filtering-based solutions (for example the Kalman filter \cite{montemerlo2002fastslam}
and the particle filter \cite{Thrun_Probabilistic_robotics}) 
and graph-based solutions \cite{burgard2009comparison, kuemmerle11icra}. 

Depending on the types of sensors used, 
one can classify the SLAM into laser-based SLAM, visual-SLAM, magnetic-SLAM, WifiSLAM, and FootSLAM. 
Laser-based SLAM uses laser-range finders to create a structural map of an environment. 
The detection of loop closure is achieved by scan matching. 
Visual-SLAM methods, utilize cameras like Kinect or Tango \cite{engelhard2011real} to construct a 3D model of the indoor scene.
Bundle adjustment \cite{Triggs00bundleadjustment} is another popular technique for SLAM that uses visual images 
and has been used in commercialized SLAM systems such as Google's Project Tango \cite{tango}. 
Magnetic-SLAM systems, exploit digital magnetic compass for localization and mapping of a device \cite{JUNG_sequence_mag_slam}. 
The loop closure is inferred by examining spatial similarity of a sequence of magnetic measurements.
For example, authors in \cite{wang_motion_slam_iros} correlated motion patterns with the magnetic field to address the SLAM problem. 
A unique magnetic fingerprint may not be guaranteed due to the distortion of the environment, 
which makes this solution challenging for real applications. 

WifiSLAM \cite{ferris2007wifi, Huang_wifi_slam_11} 
techniques use the radio signal and motion data of the device for localization and signal strength mapping in unknown environments. 
With SLAM technique, the hassle of site surveying can be avoided, and radio map can be created and updated conveniently whenever needed. 
For example, authors in \cite{ferris2007wifi} solved the WifiSLAM problem 
by mapping the high-dimensional signal strength into a two-dimensional latent space with a Gaussian process. 
Authors in \cite{Huang_wifi_slam_11} proposed a generalized and effective algorithm to solve the WifiSLAM using GraphSLAM algorithm. 
Both approaches 
assume the signal strength at two close locations are similar 
and the measurement likelihood can be modeled as a Gaussian process. 
In contrast, our approach does not require any model to describe the signal strength distribution, 
and the closeness of locations is determined by the similarity of the radio fingerprints.

FootSLAM \cite{foot_slam_original} 
uses inertial-based measurements to determine the underlying building structure. 
No ranging or visual measurement were required; 
the only features used are the probability distributions of human motions at different locations. 
Several extensions, for example ActionSlam \cite{action_slam}, 
incorporate location-orientated actions (for example entering elevators or door opening) as features 
to compensate for the IMU drifting error. 
Additional information \cite{foot_slam} can be further incorporated into FootSLAM, 
for example a prior map or signal strength from a Wi-Fi access point.
Authors in \cite{floor_plan_generation_jsac} proposed SenseWit that utilizes inertial measurements
to generate a floorplan by identifying featured locations (turning, water dispenser, and door) in indoor space.

When the indoor environment becomes huge, generating the radio map with single mobile device becomes time consuming. 
The power of crowd comes into play in this scenario. 
Mobile crowdsensing is a popular computing paradigm, which enables ubiquitous devices to collect sensing data at large scales \cite{crowd_facility_mapping,Smartphones_crowdsourcing,localisation_database_crowdsourcing}. 
This technique can be utilized to unleash the potential of mobile phones of people 
who move inside the indoor environment \cite{faragher2012opportunistic}. 
Prior research have harnessed the power of crowdsensing to reconstruct indoor floor plans 
by combining user mobility traces, images of landmarks, 
and Wi-Fi fingerprints \cite{radu2013pazl, gao2014jigsaw}.
Localization by combination of 6-DOF gyro-odometry and Wi-Fi localization has been done in \cite{jirkuu2016wifi}, using multiple robots.
Authors in \cite{cooperative_fingerprint_localization} proposed an approach to 
utilize pairwise distance measures between users to reduce the positioning error in fingerprinting-based approaches.

Our system combines crowdsensed RSS from Wi-Fi APs and dead reckoning information from a phone 
to localize a device and build a radio map of the environment. 
To improve the accuracy, 
we additionally incorporate turning features extracted from users' trajectories into our system.
Section \ref{system_overview} explains the details of the system implementation.

\section{Collaborative SLAM based on Pose Graph Optimization}
\label{system_overview}	
We present a novel approach that incorporates radio fingerprint measurement and motion information 
for collaborative SLAM in an unknown environment. 
The approach presented here does not require any prior knowledge about the map or the distribution of the access points 
nor does it need a labor-intensive phase to collect the measurements in the existing infrastructure.
Our approach features a cost-effective alternative to estimate the trajectory of multiple users in unknown environments. 
A radio map is created simultaneously, which can be used as reference to localize other users afterwards.

In our proposed collaborative framework, 
user walks in the environment and collects the radio measurements. 
Our approach merges the tracks from different users, performs loop closure detection, and optimizes the graph to generate a consistent radio fingerprint map. 
The data collected will be shared among all users through the server and each user will contribute to certain part of the map. 
The collaborative approach will accelerate the conventional way of map building. 
With our collaborative approach, the mapping of a building will become easier, 
since all users will participate in the map creation, which eliminates the expensive on-site survey phase in the conventional way of fingerprint map generation.

The goal is to estimate the entire trajectories from observations (i.e., Wi-Fi observations and motion measurements)
taken from different users at different times without a prior knowledge about the environment. 
The problem addressed here is known as SLAM, 
which has been well studied in the field of robotics \cite{RFM-SLAM, Thrun_Probabilistic_robotics, slam_trends_2016, ceres-solver}.
Among those, the graph-based approach, which formulates the problem as maximum likelihood estimation using pose graphs, 
is regarded as one of the most effective way to solve SLAM problem.
Based on the raw sensor measurements, graph-based SLAM \cite{Thrun_Probabilistic_robotics} creates a graph, 
where nodes denote the poses of the users and edges decode the constraints between two nodes.
The problem turns into graph optimization, 
which determines the best configuration of the poses by considering all constraints in the graph. 

Loop closure is important for any SLAM system 
and is considered as one of the main challenges in implementing a SLAM system in large-scale environment.
It represents a situation that users have revisited a previously observed location. 
Since the odometry will inevitably drift for long term run, 
loop closure allows to correct the accumulated odometric errors and create a consistent map of the scene.
The loop closure problem has been researched extensively using visual-based or ranging-based sensors \cite{Taketomi2017,SujiwoATNE16}, 
which are usually costly and computationally expensive. 
Instead of using the dedicated devices to perform loop closure detection, 
we focus on the radio fingerprints, 
which are available in existing indoor infrastructures and can be easily retrieved from every smartphone. 

Formally, let $\x_{1:T}^k=\{ \x_{1}^k,...,\x_{T}^k \}$ be the path of user $k$ we would like to estimate up to time $T$,
where $\x_t^k= (x_t^k,y_t^k,\theta_t^k)$ represents the global 2D location and heading of the user at time $t$. 
Let $\z_{m,i}^{n,j}$ and $\Sigma_{m,i}^{n,j}$ represent the mean and covariance of a measurement (i.e., constraint or edge) between node $\x_{i}^m$ and $\x_{j}^n$.
We use $\mathbb{C}$ to denote the set containing all pairs of constraints.
$\hat{\z}_{m,i}^{n,j}(\x_{i}^m,\x_{j}^n)$ is the prediction of a measurement based on the current configuration of node $\x_{i}^m$ and $\x_{j}^n$.
The graph-based SLAM aims to find the best configuration $\x^*$ to meet the following criteria:
\begin{equation}
\begin{split}
\x^*= \argminA_{\x} \sum_{(\x_{i}^m,\x_{j}^n) \in \mathbb{C}} & {(\z_{m,i}^{n,j}-\hat{\z}_{m,i}^{n,j}(\x_{i}^m,\x_{j}^n))}^\intercal {\Sigma_{m,i}^{n,j}}^{-1} \\
&\times(\z_{m,i}^{n,j}-\hat{\z}_{m,i}^{n,j}(\x_{i}^m,\x_{j}^n)) 
\end{split} 
\label{eq:optimization}
\end{equation}
In particular for graph-based SLAM, $\z_{m,i}^{n,j}$ is known as edge or constraint which represents a rigid-body transformation between node $\x_{i}^m$ and $\x_{j}^n$.
The transformation is a $3 \times 1$ vector which encodes the 2D translation (i.e., $x_{m,i}^{n,j}$ and $y_{m,i}^{n,j}$) and the rotation $\theta_{m,i}^{n,j}$.
The constraint can be either sequential odometry measurement (i.e., odometry-based constraint) or 
loop closure (observation-based constraint), which is determined by aligning the sensor observations at two non-consecutive poses.
Since observations are usually erroneous, all constraints are additionally parameterized with a certain degree of uncertainty (i.e., $\Sigma_{m,i}^{n,j}$). 
For laser range finders, 
the transformation $\z_{m,i}^{n,j}$ can be determined by matching two scans \cite{Lu_millos_1997} using model-based registration, for example iterative closest point (ICP). 
Given a signal strength measurement from an AP, 
it is straightforward to know if an area has been visited by a user, 
since each reported RSS value is associated with a unique MAC address. 
However, estimating the precise transformation $\z_{m,i}^{n,j}$ between two observations turns out to be tricky, 
since radio signal neither reports distance nor bearing, 
and the detection range of an AP can be up to 50 meters, 
which is usually much larger than the accumulated error of a pedestrian dead reckoning system.

The distance to an access point can be approximated by the signal strength 
via an analytical signal-to-distance model. 
This model is used by some researchers to address the SLAM problem \cite{Xiong2017ADG, ferris2007wifi, Huang_wifi_slam_11}
and indoor positioning \cite{he2016wi,Yassin_ieee_tutorials_2016}.
However, obtaining such a model is usually not practical, 
as the propagation of signal will be distorted by many environmental factors (for example, multiple path or obstruction from obstacles).
Instead of modeling them explicitly, this paper represents the location with radio fingerprint, 
which consists of the address of detected device and the measured signal strength. 
These fingerprints are location-dependent and are assumed to be unique to describe a location in an infrastructure. 
The closeness of two locations can be determined by comparing 
the degree of similarity of the fingerprints.

We claim a loop closure if the similarity between two radio measurements at $\x_{i}^m$ and $\x_{j}^n$ reaches a threshold $\vartheta_s$. 
We then infer that their locations are the same
and add a constraint $\z_{m,i}^{n,j}$, with all elements zero, to the graph.
Actually, 
the two locations are unlikely to be exactly the same, which will produce a small amount of error to the loop closure.
The error can be compensated by associating a covariance $\Sigma_{m,i}^{n,j}$ to the constraint. 
A choice of this can be a diagonal matrix by setting small values on the main diagonal. 
Our solution is a careful examination of the uncertainty of a loop based on the degree of similarity in a training phase. 
Based on training data, we automatically learn a nonparametric model to characterize the degree of similarity conditioned on the distance of two locations.

\begin{figure}
\centering
\includegraphics[width=0.49\textwidth]{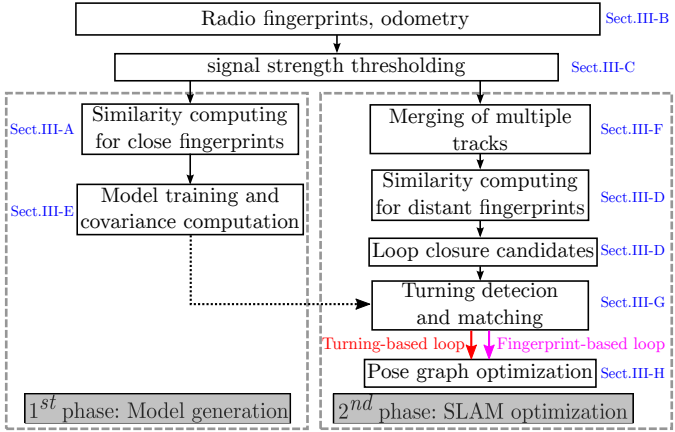}
\caption{
Illustration of our collaborative simultaneous localization and radio fingerprint mapping (C-SLAM-RF) system. 
The proposed system automatically creates a radio fingerprint map of an environment 
using the radio and odometry measurements from a group of users. 
The fingerprint-based and turning-based loops are identified and incorporated into a graph-based SLAM algorithm for optimization.
}
\label{fig:system_overview}
\end{figure}

The uncertainty of the loop closure inferred from radio fingerprints is very high, 
we therefore exploit the turning features to improve the accuracy. 
We identify the turnings using motion information, 
match the turnings, and assign the loop closure with a small covariance if a match is found. 
Figure\,\ref{fig:system_overview} gives an overview of the system.
We will describe the details of each component in our proposed solutions in the following subsections. 
We summarize the notations of the symbols used in this paper in Table\,\ref{notations}.
\begin{table}[]
\centering
\caption{Summary of important variables used in this paper}
\label{notations}
\begin{tabular}{|c|p{6.5cm}|}
\hline
Symbol & Meaning \\ \hline \hline
$\x_t^k$     & \multicolumn{1}{m{6.5cm}|}{the pose of user $k$ at time $t$, i.e., $\x_t^k=(x_t^k,y_t^k,\theta_t^k)$}    \\ \hline
$\x_{1:T}^k$     & \multicolumn{1}{m{6.5cm}|}{ the trajectories of user $k$ up to time $T$ }    \\ \hline
$\z_{m,i}^{n,j}$    & \multicolumn{1}{m{6.5cm}|}{the transformation (i.e., translation and rotation) between node $\x_{i}^m$ and $\x_{j}^n$}    \\ \hline
$\Sigma_{m,i}^{n,j}$    & \multicolumn{1}{m{6.5cm}|}{the covariance of the transformation $\z_{m,i}^{n,j}$ between node $\x_{i}^m$ and $\x_{j}^n$}  \\ \hline
$\F_t^k$     & \multicolumn{1}{m{6.5cm}|}{the radio fingerprint and pose of user $k$ recorded at time $t$, i.e., $\F_t^k=(\f_t^k,\x_t^k)$}  \\\hline
$\f_t^k$     & \multicolumn{1}{m{6.5cm}|}{the radio fingerprint at pose $\x_t^k$, which consists of the RSS from $L$ APs: $\f_t^k=\{ f_{t,1}^k,...,f_{t,L}^k \}$} \\ \hline
$\s_{m,i}^{n,j}$     & \multicolumn{1}{m{6.5cm}|}{the cosine similarity between $\f_{i}^m$ and $\f_{j}^n$}  \\ \hline
$d(\x_i^m,\x_j^n)$     & \multicolumn{1}{m{6.5cm}|}{the relative distance between pose $\x_i^m$ and $\x_j^n$ where fingerprints $\F_i^m$ and $\F_j^n$ are recorded }    \\ \hline
$\theta(\x_i^m,\x_j^n)$     & \multicolumn{1}{m{6.5cm}|}{the relative orientation between pose $\x_i^m$ and $\x_j^n$ where fingerprints $\F_i^m$ and $\F_j^n$ are recorded }    \\ \hline
$\vartheta_{s}$     & \multicolumn{1}{m{6.5cm}|}{the threshold used to claim a loop closure based on the cosine similarity $\s_{m,i}^{n,j}$} \\ \hline
$\vartheta_{r}$     & \multicolumn{1}{m{6.5cm}|}{the threshold used to filter out the low RSS measurement}  \\\hline
$r$     & \multicolumn{1}{m{6.5cm}|}{the binning size to train the model (i.e., the uncertainty given a measured similarity)} \\ \hline
$\{s_k,d_k\}_{k=1}^{K}$   & \multicolumn{1}{m{6.5cm}|}{the $K$ training samples, where $s_k$ denotes the similarity and $d_k$ denotes the distance between a fingerprint pair }    \\ \hline
$\binning(s,r)$    & \multicolumn{1}{m{6.5cm}|}{a set of samples that sits in an interval $r$ around a similarity $s$ }    \\ \hline
$var(d|s)$     & \multicolumn{1}{m{6.5cm}|}{variance of the samples given a similarity $s$ }    \\ \hline
$c(\binning(s,r))$   & \multicolumn{1}{m{6.5cm}|}{the number of samples that sits in an interval $r$ around a similarity $s$ }    \\ \hline
$a_{t}^k$   & \multicolumn{1}{m{6.5cm}|}{the acceleration measurement of user $k$ at time $t$ }    \\ \hline
$s$     & \multicolumn{1}{m{6.5cm}|}{the step length of the step counter-based pedestrian dead reckoning} \\\hline
$R_t^k(\tau)$ & \multicolumn{1}{m{6.5cm}|}{the normalized auto-correlation of the accelerometer data for lag $\tau$ at the $t$th sample of user $k$} \\
\hline
$\mu_t^k(\tau)$   & \multicolumn{1}{m{6.5cm}|}{mean of the sequential acceleration measurements for a lag $\tau$ at time $t$}    \\ \hline
$\sigma_t^k(\tau)$   & \multicolumn{1}{m{6.5cm}|}{standard deviation of the sequential acceleration measurements for a lag $\tau$ at time $t$}    \\ \hline
$c_t^k$     & \multicolumn{1}{m{6.5cm}|}{the step counter at timestamp $t$ of user $k$} \\ \hline
$w$     & \multicolumn{1}{m{6.5cm}|}{the window size to segment the inertial measurements for turning detection and matching}  \\ \hline
$\mathbb{C}_t^k$     & \multicolumn{1}{m{6.5cm}|}{the track segmentation of user $k$ at time $t$} \\ \hline
$ \overline{\mathbb{C}}_t^k$   & \multicolumn{1}{m{6.5cm}|}{the relative positions of $\mathbb{C}_t^k$ with respect to $\x_t^k$}    \\ \hline
$\theta_{t}^{k^-}$, $\theta_{t}^{k^+}$& \multicolumn{1}{m{6.5cm}|}{the mean orientation of poses with timestamps smaller or larger than $t$ in segmentation $\mathbb{C}_t^k$}
\\ \hline
$\vartheta_{f}$     & \multicolumn{1}{m{6.5cm}|}{the threshold used to claim a valid turning match}  \\ \hline
\end{tabular}
\end{table}



\subsection{Radio Fingerprints and the Similarity}
\label{sect_rf_fingerprint}
Radio fingerprinting represents location with radio signals from radio-based sensors, for example Wi-Fi APs, Bluetooth beacons, and RFID tags. 
These fingerprints are robust against location-dependent distortions as compared to the model-based approaches,
since the propagation of the radio signal in an environment is hard to predict due to the blockage of obstacles and multipath fading issue.
This is quite similar to appearance-based approach, where the scene is represented by a number of visual features.
Extracting visual features involves a large amount of computation, 
while this process can be ignored for the radio fingerprint, 
since the AP can be regarded as the unique feature for the positioning. 

We represent a fingerprint of user $k$ at time $t$ as a pair $\F_t^k={(\f_t^k,\x_t^k)}$.
$\x_t^k=(x_t^k,y_t^k,\theta_t^k)$ denotes the odometry at time $t$ when user $k$ traverses the environment.
$\f_t^k$ represents the radio measurement at location $\x_t^k$, 
which consists of the RSS values from $L$ access points: $\f_t^k=\{ f_{t,1}^k,...,f_{t,L}^k \}$.
Let $L_i^m$ and $L_j^n$ denote the number of detections in $\f_i^m$ and $\f_j^n$, respectively.
$L_{m,i}^{n,j} = \left | \f_i^m \cap \f_j^n \right |$ is used to represent the common APs in both $\f_i^m$ and $\f_j^n$. 
The similarity function $\SIM(\F_{i}^m,\F_{j}^n)$ yields a positive value, 
representing the similarity between two radio measurements, namely $\f_{i}^m$ and $\f_{j}^n$.
We apply the cosine similarity which has been extensively used in the literature \cite{RanArtur_IROS_2012} \cite{vorst2010isr}.
\begin{equation}
\s_{m,i}^{n,j}=\SIM(\F_{i}^m,\F_{j}^n)=
\frac{\sum_{l=1}^ {L_{m,i}^{n,j}} {f_{i,l}^m f_{j,l}^n}}{\sqrt{\sum_{l=1}^ {L_i^m} {\left(f_{i,l}^m\right)^{2}}}\sqrt{\sum_{l=1}^ {L_j^n} {\left(f_{j,l}^n\right)^{2}}}}
\label{eq:similarity}
\end{equation}

We refer the readers to \cite{he2016wi} \cite{vorst2011rfidta} for a comparison between different similarity measures.

\subsection{Pedestrian Dead Reckoning}
\label{pdr}
The spatial relationship between sequential poses in Equation \ref{eq:optimization} can be determined by the odometry measurements, which is known as odometry-based constraint.
Nowadays, smartphones are equipped with various types of sensors, including IMU sensor, camera, light sensor, and proximity sensor.
This enables one to implement a variety of pedestrian dead reckoning systems using different techniques.
We evaluated our system under two pedestrian dead reckoning systems (PDR): 
Tango-based PDR using visual-inertial odometry (VIO)
and step counter-based PDR using accelerometers and compass. 
The goal is to compare the approach under various tracking systems with different tracking accuracies.

Tango is developed by Google that uses visual-inertial odometry, 
to estimate the location of a device without GPS or any external referencing.
It uses visual features with a combination of inertial measurements from accelerometer and gyroscope to track the movement of a device in 3D space.
Lenovo Phab 2 Pro and Asus Zenfone AR are two examples of commercially available Tango phones.

Alternatively, 
the IMU sensor embedded inside a phone can be used to implement a step counter-based dead reckoning.
A typical IMU system is comprised of accelerometer, gyroscope, and magnetometer for motion or orientation sensing. 
Following Zee \cite{Zero_calibration}, 
we implemented the step counting based on auto-correlation.
Given the acceleration measurement $a_t^k$ of user $k$ at time $t$, 
the step counting is achieved by examining the periodic step patterns 
through normalized auto-correlation $R_t^k(\tau)$ for a lag $\tau$:
\begin{equation}
R_t^k(\tau)=
\frac{\sum_{l=0}^{l=\tau-1} \begin{bmatrix}
\begin{array}{l}
(a_{t+l}^k-\mu_t^k(\tau)) \\ 
(a_{t+l+\tau}^k-\mu_{t+\tau}^k(\tau))
\end{array}
\end{bmatrix}  }
{\tau \sigma_t^k(\tau) \sigma_{t+\tau}^k(\tau)},
\end{equation}
where $\mu_t^k(\tau)$ and $\sigma_t^k(\tau)$ are mean and standard deviation of the sequential acceleration measurements
$\{a_{t}^k,a_{t+1}^k,...,a_{t+\tau-1}^k\}$. 
The algorithm first 
identifies an optimal $\tau_{opt}$ to maximize $R_t^k(\tau)$.
Then $\tau_{opt}$ is used as a replacement of $\tau$ to count further steps.
Similar to Zee\,\cite{Zero_calibration}, 
the sampling rate of the IMU is 50Hz, 
we therefore set the initial searching window $\tau$ to $[40,100]$. 

During the walking of a person, we assume the phone is always held in front of him.
Therefore, we use the magnetometer reading to approximate the orientation of the user.
Let $c_t^k$ and $\theta_t^k$ be the step counting and the orientation of user $k$ at time $t$ respectively, 
the position (i.e., $x_t^k$ and $y_t^k$) of user $k$ is determined by:
\begin{align}
x_t^k=x_{t-1}^k+s\cdot(c_t^k-c_{t-1}^k) \cdot \cos (\theta_{t-1}^k) \\
y_t^k=y_{t-1}^k+s\cdot(c_t^k-c_{t-1}^k) \cdot \sin (\theta_{t-1}^k),
  \label{eq:pdr}
 \end{align}
 where $s$ is the step length, which is assumed to be fixed throughout the experiments. 
 The estimation of the pose in odometry frame works in recursive fashion. The initial values of $x$ and $y$ are set to zero.
 The initial headings of different PDR systems are treated differently. 
 For Tango-based PDR, the initial heading of a user is set to zero. 
 For step counter-based PDR, the heading is determined by the in-built magnetometer based on geographic cardinal directions.
 The magnetic materials in the building might affect the accuracy of the orientation estimation.
In the future, we would like to compensate for the orientation by incorporating the gyroscope readings \cite{richard_ieeesensor_2016}.

The odometry-based edge is determined based on the relative translation and rotation between the sequential odometry measurements. 
For a user $k$, 
the rigid-body transformation ($\Delta x_t^{k}$, $\Delta y_t^{k}$ and $\Delta \theta_t^{k}$) between the pose at time $t-1$ and time $t$ can be computed as:

  \begin{align}
  \begin{bmatrix} 
  \Delta x_t^{k} \\
  \Delta y_t^{k} \\ 
  \Delta \theta_t^{k} \\
 \end{bmatrix}
 =
 \begin{bmatrix} 
  \cos( \theta_{t-1}^{k} ) & -\sin( \theta_{t-1}^{k} ) & 0\\
  \sin( \theta_{t-1}^{k} ) & \cos( \theta_{t-1}^{k} ) & 0 \\
  0 & 0 & 1 \\ 
 \end{bmatrix}
 ^\top
 \begin{bmatrix}
  x_t^{k}-x_{t-1}^{k}  \\
  y_t^{k}-y_{t-1}^{k} \\ 
  \theta_t^{k}-\theta_{t-1}^{k} \\
 \end{bmatrix}
 \end{align}

\subsection{RSS Thresholding}
\label{sect_rss_threshold}
The time required to compute the similarity in Equation (\ref{eq:similarity}) grows linearly with the number of APs in the two fingerprints. 
The computational time can be significant in densely AP covered environment, which is the typical case in modern office or commercial buildings. 
A large amount of computational time can be saved if the size of the measurements can be reduced.
Therefore, we propose to filter out the RSS observation with value below a threshold $\vartheta_r$.

Thresholding prunes observations with small RSS values, 
which represent spurious readings due to multiple propagation issues in indoor environment. 
In addition, larger RSS values indicate a location close to the access point with more confidence. 
These measurements are expected to better confine the location of the user.
In the experimental section, 
we show that thresholding technique can provide a better accuracy while consuming less computational time. 
\subsection{Finding Loop Closure Candidates}
\label{sect_finding_loop_candidate}
To find the observation-based edge (i.e., constraint) between the non-consecutive poses in Equation \ref{eq:optimization}, 
we need to perform loop closure detection.
The observation-based edge consists of two different types of edges, namely similarity-based edge and turning-based edge. 
In our approach, each fingerprint carries the odometry information where the fingerprint is recorded 
(i.e., $x$, $y$, and the orientation $\theta$ of a user). 
We first compute the relative distance $d(\x_i^m,\x_j^n)$ and orientation $\theta(\x_i^m,\x_j^n)$
between the odometric poses of two fingerprints $\F_i^m$ and $\F_j^n$. 
If these values are smaller than pre-defined thresholds (50 meters and 0.3 radians for distance and orientation respectively), we compute the similarity $\s_{m,i}^{n,j}$ between them. 
We add a tuple $<\x_{i}^m,\x_{j}^n,\s_{m,i}^{n,j}>$ as a candidate of the loop closure if the similarity $\s_{m,i}^{n,j}$ exceeds a threshold $\vartheta_s$, 
which is one of the few parameters that has to be supplied by the user. 
The impact of $\vartheta_s$ on the performance is not too critical, as shown in our experiments. 
In most cases, $\vartheta_s=0.7$ gives good results. 
We reject the similarity with values smaller than $\vartheta_s$, to avoid false positive loop closures.
To improve the accuracy of the system, we further check if this loop is a turning-based loop. 
We identify a turn by examining the orientation changes and checked the fitness of their respective tracks by a matching algorithm. 
If the fitness score is higher than a threshold, we consider this loop as a turning-based loop. The detail of the detection of turning-based loop can be found in Section \ref{turning_indentification_matching}.


\begin{figure}
  \centering
    \subfigure[Experimental snapshot in our campus]{
\label{fig:environment}
    \includegraphics[height=0.25\textwidth]{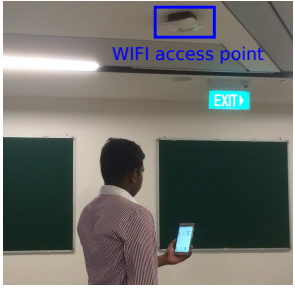}
    }   
    \hspace{0.0cm}
  \subfigure[Similarity and the distance variance in two different buildings]{
\label{fig:sim_distance}
        \includegraphics[height=0.26\textwidth]{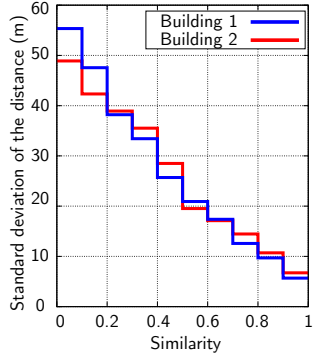}
        }
   \caption[Similarity]
{(a) Snapshot of a person holding a phone and walking in the environment;
(b) The trained variance model at two different buildings.}
\label{fig:snapshot_and_distance_variance}
\end{figure}

\begin{figure}
\centering
\includegraphics[width=0.45\textwidth]{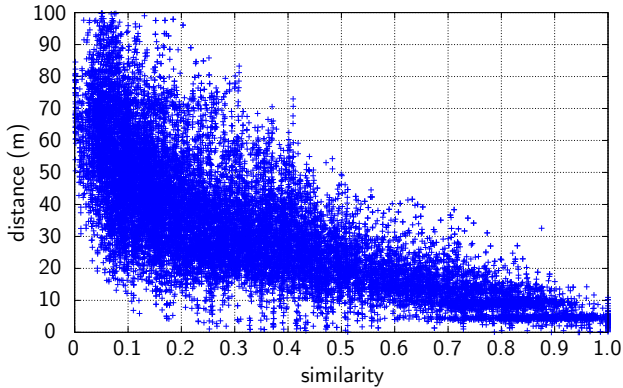}
\caption{Scatter plot of similarity versus distance from experimental data in building 1.}
\label{fig:scatter_plot}
\end{figure}

\subsection{Model Training}
\label{sect_model_training}
To optimize Equation \ref{eq:optimization}, an uncertainty estimation of the constraint is necessary for all edges in SLAM graph. 
For odometry-based edges, the parameter is obtained from the motion model. 
We now need to derive a model to represent the uncertainty of the observation-based edges.
Our solution is to train such model by passing over the observation data (i.e., odometry and radio fingerprints), 
which is recorded by the smart phone as shown in Figure \ref{fig:system_overview}. 

Our goal is to generate an uncertainty model to feature the distance variance of two radio fingerprints given their similarity. 
To build such a model, we need to know the true locations where the fingerprints are recorded or relative distance between the recorded positions, 
which is not possible without any external reference system.
Although the error from odometry accumulates in the long term, 
it is sufficiently small for a short time of duration. 
In this work, we assume odometry is accurate enough for the distance traveled less than 100 meters, 
which is suitable for most inertial tracking platforms \cite{richard_ieeesensor_2016} \cite{visual_odom_zhang_ICRA} \cite{foot_mounted}. 
For example, authors in \cite{visual_odom_zhang_ICRA} evaluated the visual odometry with wide angle and fisheye cameras, 
and showed a relative positioning error of less than 1.4\% with a distance of 538 meters traveled.
Therefore, we compute the degree of similarity for close fingerprint pairs. 
These values are annotated with the distance between the two locations using the PDR system. 
As a result, we will get a set of $K$ training samples: $\{s_k,d_k\}_{k=1}^{K}$, 
where $s_k$ is the similarity and $d_k$ is the distance of the fingerprint pair.
Figure \ref{fig:scatter_plot} shows the scatter plot of distance versus similarity in one of the buildings.
We then train a model which features the variance of distance given a similarity by binning. 
That is, for a similarity value $s$, 
we compute the variance of the samples that sites in the small interval $r$ around $s$:
\begin{equation}
var(d|s)=\frac{1}{ c (\binning (s,r)) } \sum_{k \in \binning(s,r)} { d_k}^2
\end{equation}
where $c(\binning)$ counts the number of samples in interval $r$. 
$var(d|s)$ denotes the variance of the distance $d$ given a similarity $s$. 
Although binning is a simple way for smoothing, 
the computation is efficient, 
since assigning the sample into a bin is straightforward. 
One example of the variance computed in two different buildings is shown in Figure \ref{fig:sim_distance}. 
The resulted variance is stored in a look up table, 
which could be used in the second stage of SLAM, as shown in Figure \ref{fig:system_overview}. 

\subsection{Merging Tracks at Different Times}
\label{track_merge}
To leverage the power of crowdsourcing, we utilize the measurements captured from different users to generate a radio map of the environment. 
This involves the loop closure detection between different users in order to correct their paths using the power of crowdsourcing. 
The tracks recorded from different users are based on different reference systems, for example different starting positions. 
The determination of orientation is different for the two pedestrian dead reckoning systems.
The Tango-based PDR estimates the orientation by visual-inertial odometry based on the starting pose and 
the step counter-based PDR determines the orientation based on the compass readings and is not relevant to any starting position. 
Therefore, these trajectories are needed to be merged into the same coordinate system to guarantee a robust loop closure detection between different users.

In this paper, we start the tracking by assuming all users passing by the same place. 
This is quite reasonable since users may pass through several key landmarks in an environment for example entrance of a building or the elevator. 
One might argue that in large buildings for example airport, 
not all users share the common place. 
It is possible to first build several sub-maps, and then merge them into a large and consistent map \cite{multi_SLAM_lidar} \cite{merge_multiple_map} \cite{Optimization_2Dmap}. 
An edge is added to connect the first nodes in different tracks. 
For the transformation matrix $\z_{m,i}^{n,j}$,
we set $x_{m,i}^{n,j}$ and $y_{m,i}^{n,j}$ to zero 
and the covariance is obtained by checking the covariance table as detailed in Section \ref{sect_model_training}.
Due to the omnidirectional characteristics of antennas, the facing of a user has very little impact on the radio signals, 
therefore the radio fingerprint does not deliver any orientation information. 
This is the reason why Section \ref{sect_model_training} does not model orientation variance with respect to the similarity.
Based on two fingerprint observations, we have no knowledge about how accurate is the relative orientation between two poses, 
we therefore set $\theta_{m,i}^{n,j}$ to zero and give a very large covariance value (i.e., 1000) to the edge, 
meaning that we are not able to infer the relative angle from two radio fingerprint observations. 
For users starting from arbitrary locations, 
we refer to \cite{luo_localization} \cite{Zhou_activity_landmark} \cite{Shen_walkie} \cite{Li_crowdsourcing_traces}
to merge the paths based on the radio measurements and activity landmarks in our future work.

\subsection{Turning Identification and Matching}
\label{turning_indentification_matching}
A typical indoor infrastructure often contains various landmarks, 
such as turnings, elevators, and staircases. 
These landmarks are unique in an existing infrastructure 
and can be used as a good feature for loop closure identification in a SLAM system. 
In this case, a loop closure can be claimed 
if two landmarks match each other.
Due to the physical constraint of an environment, 
such kind of loop closure provides lower positioning uncertainty as compared to fingerprinting-based loops. 
For example, the size of an elevator is usually less than 3 meters, 
and the turning radius during human walking is smaller than a typical corridor width (i.e., 5 meters), while the positioning error using fingerprinting-based approach is usually larger than 5 meters.

Here we only focus on the turning features in a trajectory, 
which is regarded as one of the most common indoor landmarks.
A slide window is used to produce segmentations of the track.
In particular, we define a segmentation $\mathbb{C}_t^k$ of user $k$ at time $t$ 
as a collection of sequential poses with a window size of $w$, 
i.e., $\mathbb{C}_t^k=\{\x_{t'}^k\,|\,-\frac{w}{2}  \le t'-t \le \frac{w}{2} \}$.
For each loop closure candidate $<\x_{i}^m,\x_{j}^n>$, we check if there are turnings 
at these poses by examining the orientation change in segment $\mathbb{C}_i^m$ and $\mathbb{C}_j^n$.
If yes,
we try to match the two segmentations $\mathbb{C}_i^m$ and $\mathbb{C}_j^n$ using ICP (Iterative Closest Point) \cite{icp_original}.
If the fitness score (i.e., average of squared distances between the correspondence points)
between $\mathbb{C}_i^m$ and $\mathbb{C}_j^n$ is smaller 
than a predefined threshold $\vartheta_f$, we regard loop $<\x_{i}^m,\x_{j}^n>$ as the turning-based loop. 
Otherwise, this loop closure is referred to as fingerprint-based loop, as shown in Figure \ref{fig:system_overview}. 
The transformation $\z_{m,i}^{n,j}$ of both types of loop is set to zero.
We treat the covariance matrix $\Sigma_{m,i}^{n,j}$ differently:
for fingerprint-based loop, 
the covariance along $x$ and $y$ can be found at the looking up table computed previously in Section \ref{sect_model_training};
for turning-based loop, 
we set the covariance along $x$ and $y$ to 5.0, 
which is smaller than fingerprint-based loop (i.e., 8.0 in Figure \ref{fig:environment} with the highest fingerprint similarity). 
The orientation covariance in $\Sigma_{m,i}^{n,j}$ is set to 1000, 
meaning that we do not have any knowledge about the orientation of the two poses 
by the radio observations or the matching of two turnings. 
The details of turning identification and matching will be described in subsequent parts:

\begin{figure}
\centering
\includegraphics[width=0.45\textwidth]{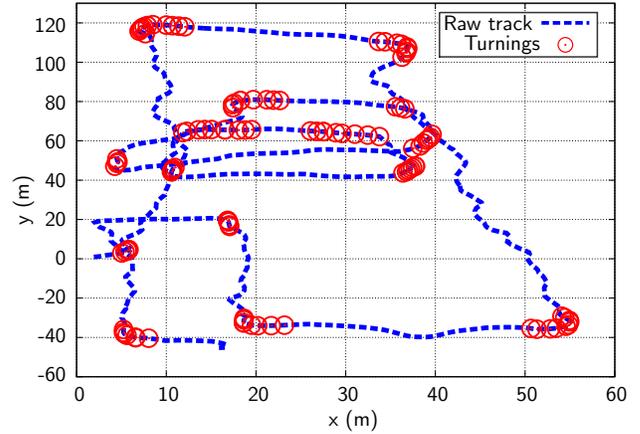}
\caption{Example of the track (blue color) and turnings detected (red circles) with a window size of $20$ based on the approach described in Section \ref{turning_indentification_matching}.
In total, 111 turnings are identified. These turnings are further examined by turning matching module to find the potential turning-based loops.}
\label{fig:turning_detection}
\end{figure}

\begin{figure}
\centering
\includegraphics[width=0.49\textwidth]{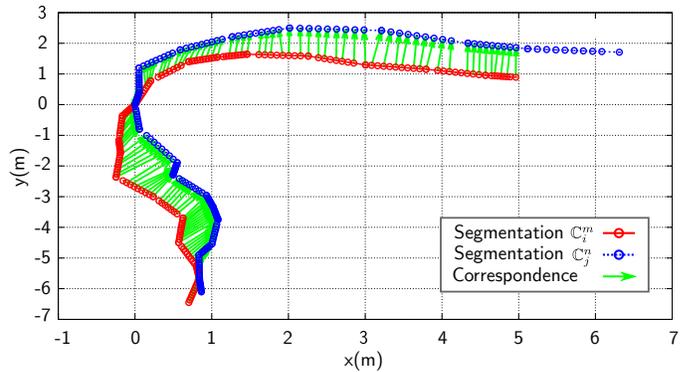}
\caption{Example of two track segmentations (i.e., $\mathbb{C}_i^m$ and $\mathbb{C}_j^n$ in blue and red color, respectively), 
and the correspondence points found using ICP (green color). 
A fitness score is calculated to determine if this candidate is a turning-based loop.}
\label{fig:turning_matching}
\end{figure}

\subsubsection{Turning Identification}
\label{turning_indentification}
We first segment the compass data and find the potential turnings \cite{turn_detection_trans}.
For the pose $\x_t^k$, we calculate the mean orientations 
for the poses with timestamps smaller and larger than $t$ 
in the segmentation $\mathbb{C}_t^k$, i.e.,
$\theta_{t}^{k^-}= \mu \{\theta_{t'}^k | t'-t \le 0\}$ and 
$\theta_{t}^{k^+}= \mu \{\theta_{t'}^k | t'-t \ge 0\}$. 
If $|\theta_{t}^{k^+} - \theta_{t}^{k^-}|$ is higher than a threshold 
(for example $\frac{\pi}{3}$ as suggested in \cite{turn_detection_trans}), a turning is identified.
The window size $w$ here has the impact on the performance of turning detection 
and we show its impact on the accuracy in the experimental section.
One example of the track and turnings detected are shown in Figure \ref{fig:turning_detection}. 
A better approach to improve the accuracy of turning detection can be found in \cite{sun_corner_detection_2017}.

\subsubsection{Turning Matching}
\label{turning_match}
The ICP aims to find a transformation (translation and rotation) between two point clouds 
that minimizes the sum of the square distance between the correspondence points. 
This approach has been extensively used to match 2D laser scans in the field of robotics. 

To find the correct transformation using ICP, 
an appropriate initial transformation has to be provided,
otherwise ICP will fall into the local minimum. 
Rather than using the global raw odometry for a segmentation $\mathbb{C}_t^k$, 
we use the relative translation between $\x_{t'}^k$ and $\x_t^k$,
i.e., $\mathbb{\overline{C}}_t=\{ \mathbf{T}_{\x_{t}^k}^{\x_{t'}^k} \,|\,-\frac{w}{2} \le t'-t \le \frac{w}{2} \}$.
Since the sampling rate of our pedestrian tracking system 
is too low (less than 1.0 HZ), 
we further interpolate the trajectory to get a large amount of locations for performing ICP.
An illustration of the turning matching using ICP is shown in Figure \ref{fig:turning_matching}. 

We finally considers $<\x_{i}^m,\x_{j}^n>$ as a valid match (i.e., turning-based loop)
if the fitness score is smaller than a threshold $\vartheta_f$. 
Robust loop closure detection is essential to a SLAM system,  
as incorrect loop closures will ruin the consistency of trajectory and the map. 
Other heuristic approach can be applied to further examine the loops and filter out the suspicious ones. 
Authors in \cite{GalvezIROS11}, for example, 
proposed an approach to group the close loop closures 
and check the temporal and spatial consistency for robust loop closure detection.
However, this technique goes beyond the scope of this paper, 
hence, we add the loop closures without performing futher consistency check. 
\subsection{Pose Graph Optimization}
\label{pose_graph_optimization}
The Equation \ref{eq:optimization}, which represents a graph consisting of poses (i.e., nodes) and constraints (i.e., edges), 
is finally optimized through the pose graph optimization algorithm GraphSLAM.
For the implementation, 
we choose Levenberg-Marquardt in g2o \cite{kuemmerle11icra}, 
which is freely available and is one of the state-of-the-art SLAM algorithms\footnote{https://github.com/RainerKuemmerle/g2o}. 

\begin{figure}
  \centering 
  \subfigure[Evaluation of Tango-based pedestrian dead reckoning system]{
\label{fig:trajectory}
        \includegraphics[width=0.49\textwidth]{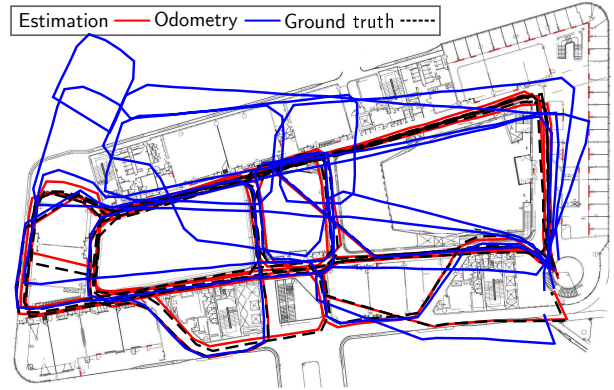}
        }        
        \subfigure[Evaluation of step counter-based pedestrian dead reckoning system]{
\label{fig:trajectory}
        \includegraphics[width=0.49\textwidth]{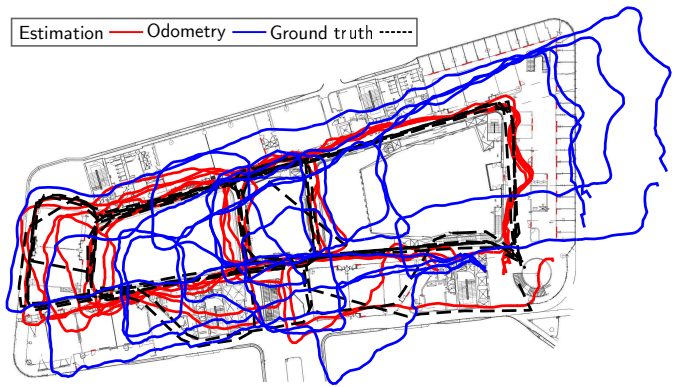}
        }    
   \caption[Floor plan.]
{Ground truth, odometry, and estimated path with our proposed approach under two different pedestrian dead reckoning (PDR) systems, namely Tango-based PDR and step counter-based PDR.}
\label{fig:floor_plan}
\end{figure}

\begin{figure}
\centering
\includegraphics[width=0.49\textwidth]{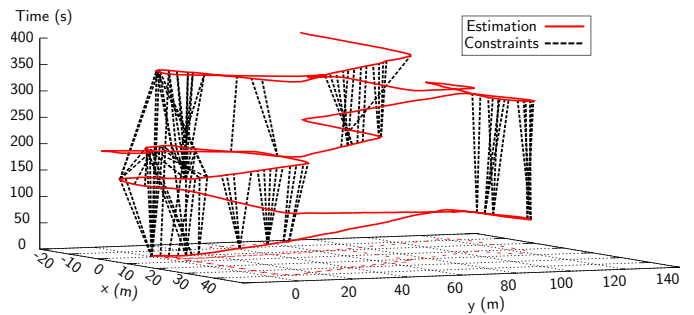}
\caption{
Part of the estimated track over time using Tango-based pedestrian dead reckoning and fingerprint-based constraints inferred with a similarity threshold of $\vartheta_s=0.7$. 
}
\label{fig:constraints}
\end{figure}

\begin{figure*}
  \centering 
       \subfigure[Visualization of track from user1 using Tango-based PDR]{
\label{fig:trajectory}
        \includegraphics[height=0.38\textwidth]{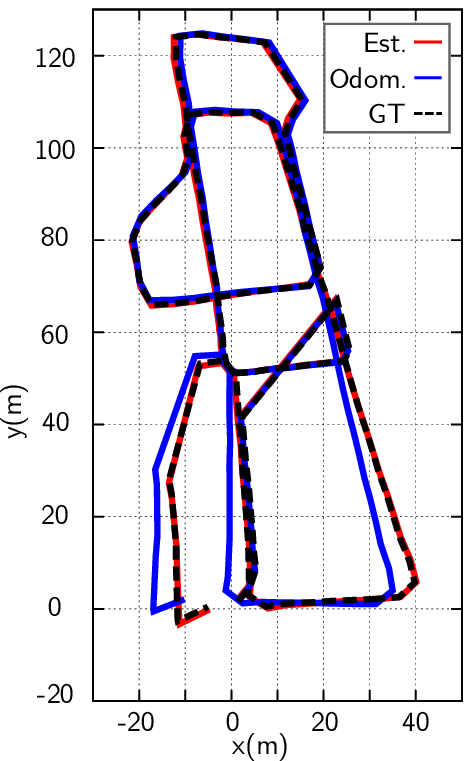}
        }
        \subfigure[Visualization of track from user2 using Tango-based PDR]{
\label{fig:trajectory}
        \includegraphics[height=0.38\textwidth]{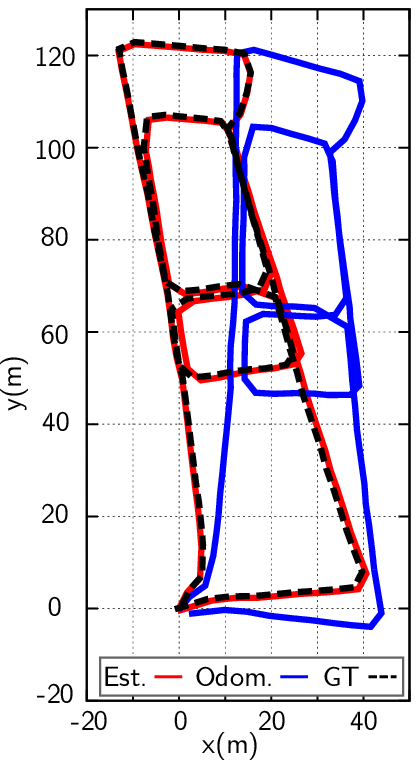}
        }        
 \subfigure[Visualization of track from user3 using Tango-based PDR]{
\label{fig:tracking_error}
        \includegraphics[height=0.38\textwidth]{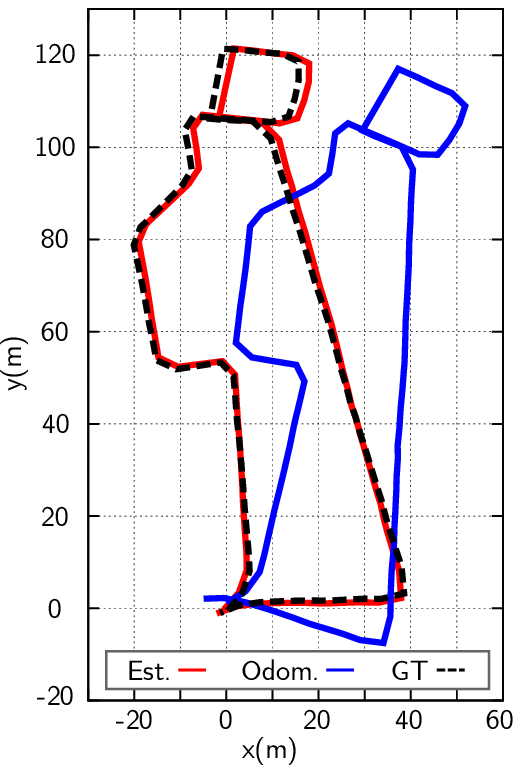}
        }        
        \subfigure[Visualization of track from user4 using Tango-based PDR]{
\label{fig:impact_of_different_imu_noise}
    \includegraphics[height=0.38\textwidth]{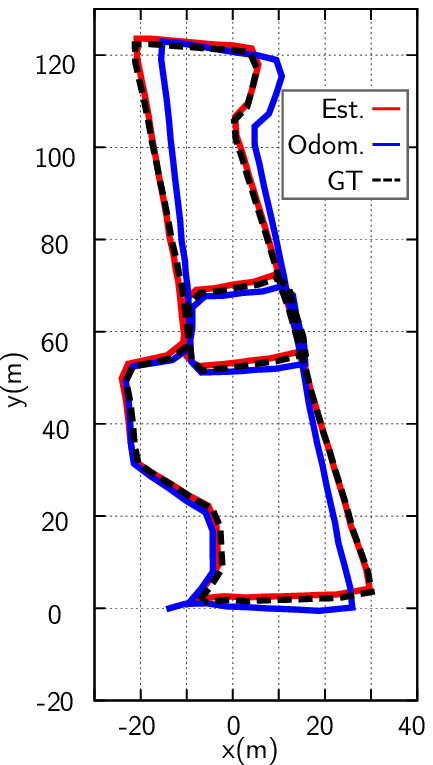}
    }
      \subfigure[Visualization of track from user1 using step counter-based PDR]{
\label{fig:trajectory}
        \includegraphics[height=0.36\textwidth]{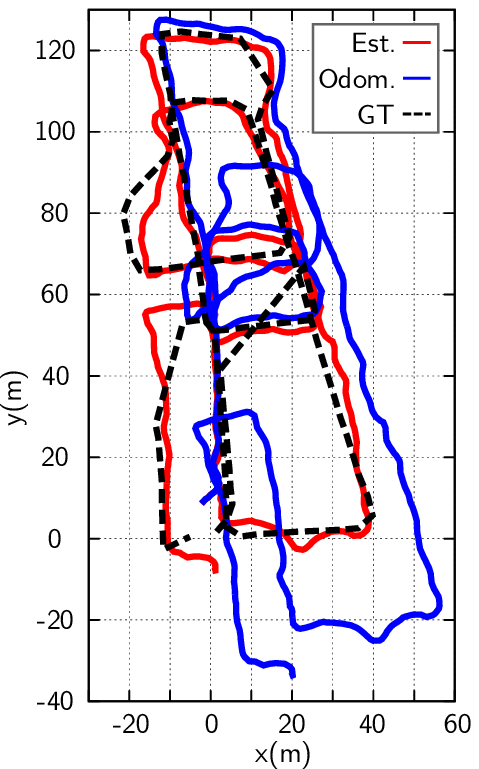}
        }        
        \subfigure[Visualization of track from user2 using step counter-based PDR]{
\label{fig:trajectory}
        \includegraphics[height=0.36\textwidth]{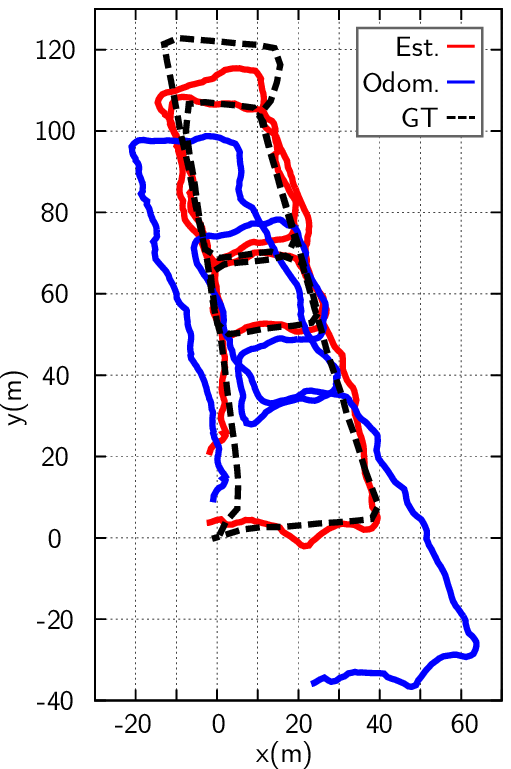}
        }
        \subfigure[Visualization of track from user3 using step counter-based PDR]{
\label{fig:trajectory}
        \includegraphics[height=0.36\textwidth]{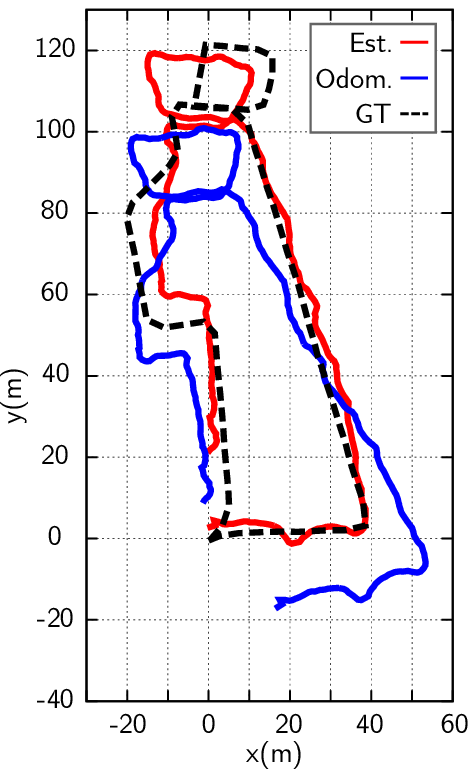}
        }
        \subfigure[Visualization of track from user4 using step counter-based PDR]{
\label{fig:trajectory}
        \includegraphics[height=0.36\textwidth]{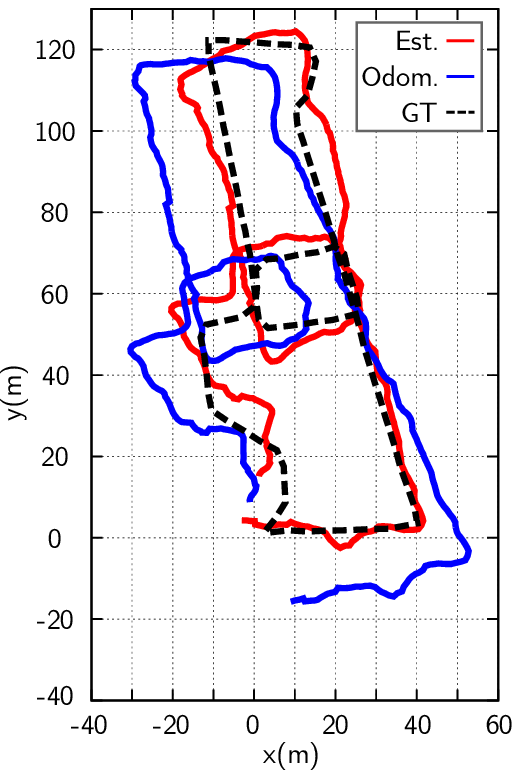}
        }
   \caption[Tracking of fusion.]
{A comparison between ground truth from point cloud-based SLAM, estimated track using our approach, 
and raw odometry of the three individual users using two different pedestrian dead reckoning systems (i.e., Tango and step counter).
}
\label{fig:example_track}
\end{figure*}

\begin{table*}[]
\centering
\caption{the role of parameters in the proposed method and guidelines on how to choose their values.}
\label{important_parameters}
\begin{tabular}{|c|c|c|c|}
\hline
Parameter & Range  & Guidelines to choose the parameters \\ \hline \hline
$\vartheta_{s}$     & [0, 1] & \multicolumn{1}{m{13.0cm}|}{Threshold used to claim a loop closure based on similarity between two radio fingerprints. 
A large $\vartheta_{s}$ will produce less loop closures and a small $\vartheta_{s}$ will result in false loop closures. 
A setting of $\vartheta_{s}=0.7$ is recommended in our approach.}    \\ \hline

$\vartheta_{r}$     & [-90, -50] & \multicolumn{1}{m{13.0cm}|}{Threshold used to filter out the low RSS measurements. The choice of $\vartheta_{r}$ depends on the distribution of the 
received signal strength in a radio environment. A value of $\vartheta_{r}=-70$ is suggested in our approach.}    \\ \hline

$r$     & [0, 1] & \multicolumn{1}{m{13cm}|}{Binning size to train the variance model. 
A small $r$ requires more training time and training samples.
A large $r$ is not able to characterize the detail of the variance model. 
A setting of $r=0.2$ produces the best results in our approach.}    \\ \hline

$w$     & [0, 200] & \multicolumn{1}{m{13cm}|}{Window size to identify a turn. 
A large $w$ will result in more false turnings and a small $w$ is not able to infer the correct turnings. $w=40$ is suggested in our approach. }    \\ \hline

$\vartheta_{f}$     & [0, 1] & \multicolumn{1}{m{13.0cm}|}{Threshold used to claim a valid turning match. 
A large $\vartheta_{f}$ will discard the true turning loop candidates and a small $\vartheta_{f}$ will incorporate too many false turning loops.
$\vartheta_{f}=0.5$ is appropriate for the application.
}    \\ \hline

\end{tabular}
\end{table*}

\begin{table*}[]
\centering
\caption{ Evaluation of the system performance with respect to different RSS thresholds $\vartheta_r$. 
The table shows the positioning accuracy (mean, standard deviation, maximum, and median) in meters, 
average number of mac addresses detected, and average computational time for loop closure detection.
}
\label{table_rss}
\begin{tabular}{|c|c|c|c|c|c|c|c|c|}
\hline
\multirow{2}{*}{\begin{tabular}[c]{@{}c@{}} \\ $\vartheta_r$ \end{tabular}} & \multicolumn{4}{c|}{Tango-based}              & \multicolumn{4}{c|}{Step counter-based} \\ \cline{2-9} 
                       & \begin{tabular}[c]{@{}c@{}}Mean$\pm$ \\ Std. Dev.\end{tabular} & Max./Median & \begin{tabular}[c]{@{}c@{}}Number \\ of MAC\end{tabular} & \begin{tabular}[c]{@{}c@{}}Comput. \\ time (s)\end{tabular} 
                       & \begin{tabular}[c]{@{}c@{}}Mean$\pm$ \\ Std. Dev.\end{tabular} & Max./Median & \begin{tabular}[c]{@{}c@{}}Number \\ of MAC\end{tabular} &\begin{tabular}[c]{@{}c@{}}Comput. \\ time (s)\end{tabular} \\ \hline
                 -90  & 1.39$\pm$1.51    &  7.86/1.37   &  179.23   &  24.50    &   14.95$\pm$7.09               &    37.72/14.85  & 129.36    & 426.37         \\ \hline
                  -80  &  1.26$\pm$1.21 &  6.30/1.19   &  137.81   &  19.53   &    5.91$\pm$3.76             &  20.87/5.14     &  70.08   &  212.07   \\ \hline
                  -75    & 1.06$\pm$0.49    &  4.53/0.94   & 81.29   &  12.31 &  5.40$\pm$3.60               &     21.52/4.56  &  43.36   &  127.67       \\ \hline
                     -70 & 0.77$\pm$0.46   &  3.79/0.73   &  47.97   &  5.73 &  5.19$\pm$3.48                &    21.64/4.47    & 25.49    &  72.34      \\ \hline
                      -65&  0.92$\pm$0.44    &  3.87/0.84   &  28.94   & 3.29 &     5.43$\pm$3.43             &  20.40/4.70    & 13.41    &  37.78      \\ \hline
                       -60&  1.24$\pm$0.89    & 4.91/1.13   &  16.22   &  1.76 &   5.95$\pm$3.82              & 22.78/5.06      &  6.08   &  20.36      \\ \hline
                       -50& 2.44$\pm$0.95  & 5.81/2.12  &  3.29   &  0.37&    11.51$\pm$5.82         &   35.03/11.52   & 0.67    &  3.08   \\ \hline
                       odom.& 8.03$\pm$7.80   &   31.55/4.22   &  NA   &  NA &    15.39$\pm$7.89         & 41.23/14.58   & NA    &  NA   \\ \hline

\end{tabular}
\end{table*}

\section{Experimental Results}
\label{experimental_evaluations}
\subsection{Experimental Details}
\label{Implementation Detail}
We program two smart phones (Lenovo Phab 2 Pro with Android 6.0.1 and Sony Xperia Z3 with Android 5.0.2) 
to receive the signal strength from APs and perform pedestrian dead reckoning.
In particular, 
the Lenovo phone uses the Tango for position tracking and 
the motion tracking data is recorded every five seconds due to its high tracking accuracy. 
We implement the step counting on a Sony phone and record the step counting and the compass readings every one second.
To evaluate the performance of the proposed approach, 
we conducted experiments on the Level 3 of Building 2 at Singapore University of Technology and Design 
with a size of 130m$ \times $70m (see Fig. \ref{fig:floor_plan}).
This environment is comprised of corridors, concrete walls, and wide open space. 
We asked a person to hold two phones (i.e., Lenovo Phab 2 Pro and Sony) 
and walk in the building along different paths with a regular walking speed. 
Four tracks were recorded at different times to show the power of the collaborative SLAM. 
For each track, the user started from the same position. 
The total distance traveled is 1906 meters with a duration of 2179 seconds and a number of 1702 unique MAC addresses are detected.
This results in four log files with a duration of 687, 582, 439, and 471 seconds respectively. 
The step length $s$ is fixed to 0.7m throughout this paper. 
Fig. \ref{fig:environment} shows a snapshot during the experiment.

\subsection{Ground Truth and Accuracy Comparisons}
\label{ground_truth}
To extract the ground truth as comparison, 
we optimized the track from Tango using GraphSLAM taken 3D point clouds as input. 
We implement loop closure detection based on the point cloud library (PCL) \cite{pcl_icra_2011}.
We identify Harris keypoints in a pair of point clouds and compute the corresponding SHOT 
(Signature of Histograms of OrienTations) descriptors \cite{shot_feature}. 
We match these descriptors with k-nearest neighbors algorithm (k-NN) and find an initial transformation using SVD (singular value decomposition).
The transformation is further refined by ICP.
If the number of matched points exceeds a threshold (half size of the point cloud),
a loop closure is confirmed and added to the graph as constraints. 
We treat this optimized path as the ground truth to evaluate the accuracy of our system. 

We show the accuracy by the root mean square error (RMSE) between the ground truth and the estimation.
Our experiments show that we are able to achieve an accuracy of 0.6 meters with Tango-based PDR 
and 4.76 meters of a step counter-based PDR with a size of 130m$ \times $70m, as shown in Figure \ref{fig:floor_plan}.
The optimized track is annotated with the radio measurement 
and can serve as the radio map for the positioning of another user. 
In Table \ref{important_parameters}, we show the important parameters used in this paper and the remarks of how to choose these parameters. 
The final positioning error is calculated at the end of the process after the loop closure detection, 
turning matching, and pose graph optimization. Larger positioning errors are expected in a real time system, 
since we have to process the incoming data in a sequential way and provide pose estimation at regular intervals before loop closures are detected.

To implement a practical indoor localization system, the mechanism to deal with the change of the radio environment is a necessity. 
The evolution of the radio environment (for example adding or removing the access points)
can be examined by looking at the signal variance at similar locations, 
as shown in \cite{adaptive_fingerprint} \cite{self_updating_radio_maps} \cite{crowdsourced_signal_map} \cite{automated_construction} \cite{PF_radiomap_recalibrate}.
Some MAC addresses might be static for a fixed duration of time, but are essentially mobile for example personal hotspots. 
A good way to address the mobile hotspots issue is to filter out the MAC addresses by the organizationally unique identifier (OUI), 
which is used to uniquely identify a vendor\footnote{http://standards-oui.ieee.org/oui.txt}. 
The MAC address from a phone manufacturer should be removed from the detection list to prevent the uncertainty of incorporating the additional mobile hotspot observations.
Another approach to filter out the mobile hotspots is to look at the spatial relations of the detected positions of a particular MAC address \cite{WOLoc} \cite{learning_wifi_hotspots} \cite{redundant_access_point_reduction}. 

\subsection{Impact of Different RSS Threshold $\vartheta_r$}
\label{evaluation_rss_threshold}
We examined the influence of RSS thresholding on the accuracy in this series of experiments. 
We set the similarity threshold $\vartheta_s=0.7$ and use a binning size $r=0.2$. 
We chose $\vartheta_r$ values between -90 and -50 to evaluate the mean accuracy, 
as listed in Table \ref{table_rss}. 
As compared to the raw odometry, 
our approach can effectively reduce the accumulated odometry error:
with the setting of $\vartheta_r=-70$,
our approach improves the positioning accuracy by 
90.4\% and 66.3\% for Tango (from 8.03m to 0.77m) and step counter (from 15.39m to 5.19m) respectively.
In addition, 
the accuracy of the PDR has a very high impact on the accuracy achieved with our SLAM system:
Tango shows a good motion tracking performance 
and we achieved an accuracy of 0.77m with a threshold $\vartheta_r=-70$. 
The accuracy achieved with Tango is better than state-of-the-art fingerprinting-based approaches \cite{ran_ieee_sensors2017, Yassin_ieee_tutorials_2016}. 
While the step counter results in a large amount of accumulated odometry error 
and the accuracy obtained with our approach is worse (5.19m). 
A further investigation to the PDR system will help to improve the accuracy, 
which will be one of our future work.
One has to note that 
the accuracy is achieved without training 
as opposed to the fingerprinting-based approaches, 
where a time-consuming phase to collect and annotate the fingerprints is prerequisite to guarantee a good positioning accuracy.

Table \ref{table_rss} also shows that we maintain a good accuracy with an RSS threshold between -75 and -65, 
while the computational time decreases considerably with the thresholding technique. 
As an example, 
for Tango-based system, 
a threshold of -70 reduces the computation time to 5.73 seconds 
as compared to a threshold of -90 (i.e., 24.28 seconds). 
At the same time, the accuracy even increases by 0.62 meters (error drops from 1.39m to 0.77m).
A suitable threshold produces a good accuracy, as it will filter out the suspicious radio signals. 
However, a threshold larger than -65 leads to a bad result 
(for example, 1.24 meters of accuracy with $\vartheta_r=-60$ for Tango-based system). 
The ground truth, estimation, and odometry of individual tracks 
using two different pedestrian dead reckoning systems are visualized in Figure \ref{fig:example_track}.
A part of estimated trajectory and the constraints inferred are shown in Figure \ref{fig:constraints}.

\begin{table}[]
\centering
\caption{
Evaluation of the proposed approach with two different pedestrian dead reckoning systems under different settings of similarity threshold $\vartheta_s$.
The table shows the positioning accuracy (mean and standard deviation) in meters and the number of constraints inferred.
}
\label{table_sim}
\begin{tabular}{|c|c|c|c|c|}
\hline
\multirow{2}{*}{\begin{tabular}[c]{@{}c@{}} \\ $\vartheta_s$ \end{tabular}} & \multicolumn{2}{c|}{Tango-based}                                         & \multicolumn{2}{c|}{Step counter-based}                                           \\ \cline{2-5} 
                           & \multicolumn{1}{c|}{\begin{tabular}[c]{@{}c@{}}Mean $\pm$\\Std. Dev\end{tabular}} & \multicolumn{1}{c|}{\begin{tabular}[c]{@{}c@{}}No. of\\constr.\end{tabular}}
                       & \multicolumn{1}{c|}{\begin{tabular}[c]{@{}c@{}}Mean $\pm$\\Std. Dev\end{tabular}} & \multicolumn{1}{c|}{\begin{tabular}[c]{@{}c@{}}No. of\\constr.\end{tabular}} \\ \hline
		      0.95 &5.63$\pm$6.76 &18 &    13.86$\pm$7.67                                              &    2098                 \\ \hline
                      0.9 &2.15$\pm$0.71&94 &    6.29$\pm$3.67                                                    &  4241              \\ \hline
                       0.8 &1.06$\pm$0.51&384 &   5.49$\pm$3.61                                                     &   10582            \\ \hline
                       0.7 &0.77$\pm$0.46&715 & 5.19$\pm$3.48                                                       &   17020           \\ \hline
                       0.6 &0.81$\pm$0.76& 1150&      5.35$\pm$3.34                                                  &23859               \\ \hline
                       0.4 &0.80$\pm$0.79&2198 &    5.49$\pm$3.66                                                    &    42625           \\ \hline
                       0.2&1.06$\pm$0.89&3494 &    5.94$\pm$4.02                                                    &    66423         \\ \hline                       
                       0.1&1.09$\pm$0.95& 4517&    6.09$\pm$3.94                                                    &    85834 \\ \hline                       
\end{tabular}
\end{table}

\begin{table}[]
\centering
\caption{
Evaluation of the mean positioning accuracy under the impact of different settings of $\vartheta_s$ and $\vartheta_r$ for Tango-based approach.
}
\label{table:positioning_under_theta_s_theta_r}
\begin{tabular}{|c|c|c|c|c|c|c|c|}
\hline
\multirow{2}{*}{$\vartheta_s$} & \multicolumn{7}{c|}{$\vartheta_r$}  \\ \cline{2-8} 
                                          & -90  & -80  & -75  & -70  & -65  & -60  & -50  \\ \hline
0.9                                       & 2.96 & 2.65 & 2.38 & 2.15 & 1.81 & 2.08 & 2.61\\ \hline
0.8                                       & 1.68 & 1.29 & 1.16 & 1.06 & 1.23 & 1.41 & 2.58\\ \hline
0.7                                       & 1.39 & 1.26 & 1.06 & 0.77 & 0.92 & 1.24 & 2.44\\ \hline
0.6                                       & 1.22 & 1.02 & 0.87 & 0.81 & 0.89 & 1.27 & 2.35\\ \hline
0.5                                       & 0.88 & 0.83 & 0.78 & 0.82 & 0.93 & 1.57 & 2.33\\ \hline
0.4                                       & 0.99 & 0.91 & 0.82 & 0.80 & 0.85 & 1.54 & 2.40\\ \hline
0.3                                       & 1.19 & 1.16 & 1.08 & 0.92 & 0.89 & 1.61 & 2.35\\ \hline
0.2                                       & 1.26 & 1.27 & 1.22 & 1.06 & 0.96 & 1.65 & 2.35\\ \hline
0.1                                       & 1.28 & 1.29 & 1.23 & 1.09 & 1.01 & 1.70 & 2.30 \\ \hline
\end{tabular}
\end{table}
\subsection{Impact of Different Similarity Threshold $\vartheta_s$}
\label{evaluation_rss_threshold}
Next, we performed a series of experiments to 
examine the influence of accuracy with respect to different similarity thresholds $\vartheta_r$. 
We show the results in Table \ref{table_sim}. 
We fixed RSS threshold $\vartheta_r=-70$ and a binning size $r=0.2$. 
We increased the similarity threshold $\vartheta_s$ from 0.1 to 0.95 to 
evaluate the accuracy and the number of constraints inferred.
From Table \ref{table_sim}, 
we can observe that the number of constraints is different for Tango and step counter-based PDR due to different sampling rates of the device 
(5 seconds for Tango and 1 second for step counter): 
Tango-based PDR offers a small number of constraints as compared to step counter-based PDR. 
In addition, the threshold has a high impact on the accuracy and the number of constraints. 
Applying a high threshold will result in a small number of constraints and a decrease of the accuracy.
For Tango-based system, 
we obtain a mean accuracy of 0.77m with $\vartheta_s=0.7$, 
which is an improvement of 86.3\% as compared to the mean accuracy of 5.63m with $\vartheta_s=0.95$.
Yet, such an improvement is at the expense of a higher number of constraints added 
(i.e., 715 constraints with $\vartheta_s=0.7$ as compared to 18 with $\vartheta_s=0.95$).
But the accuracy does not get improved with a threshold smaller than 0.6. 
One reason could be because a low similarity value 
will always come along with a very large covariance,
and has very less strength to correct the odometric error. 
A setting of $\vartheta_s=0.7$ seems to be a good trade off between the accuracy and the number of constraints inferred.
Table \ref{table:positioning_under_theta_s_theta_r} showed the accuracy by jointly optimizing the parameters $\vartheta_s$ and $\vartheta_r$ for Tango-based approach. 
As can be seen from this table, a careful examination of $\vartheta_s$ and $\vartheta_r$ will improve the accuracy. 
A too large or too small will obviously deteriorate the performance of our approach.


\begin{table}[]
\centering
\caption{
Evaluation of the positioning accuracy (mean, standard deviation, median, and maximum in meters) with two pedestrian dead reckoning systems under the impact of different configurations of binning size $r$.
}
\label{table_binning}
\begin{tabular}{|c|c|c|c|c|}
\hline
\multirow{2}{*}{\begin{tabular}[c]{@{}c@{}} \\ $r$ \end{tabular}} & \multicolumn{2}{c|}{Tango-based} & \multicolumn{2}{c|}{Step counter-based}                                           \\ \cline{2-5} 
                       & \multicolumn{1}{c|}{\begin{tabular}[c]{@{}c@{}}Mean $\pm$\\Std. Dev\end{tabular}} & \multicolumn{1}{c|}{\begin{tabular}[c]{@{}c@{}}Max. /\\Median\end{tabular}}
                       & \multicolumn{1}{c|}{\begin{tabular}[c]{@{}c@{}}Mean $\pm$\\Std. Dev\end{tabular}} & \multicolumn{1}{c|}{\begin{tabular}[c]{@{}c@{}}Max. /\\Median\end{tabular}} \\ \hline
                   1.0  & 1.03$\pm$0.63&4.29/0.89	&     6.23$\pm$3.84                                                &      24.83/5.59 \\ \hline
                   0.8  & 0.85$\pm$0.50 &4.22/0.85	&    5.71$\pm$3.50                                                    &     21.97/5.03 \\ \hline
                   0.6  &0.84$\pm$0.39 &3.62/0.79	&     5.32$\pm$3.54                                                   &    21.78/4.60 \\ \hline
                   0.4  &0.81$\pm$0.62 &4.26/0.76	&  5.22$\pm$3.51                                                   &   21.93/4.46 \\ \hline
                   0.2  & 0.77$\pm$0.46 & 3.79/0.73	& 5.19$\pm$3.48                                                 &21.64/4.47  \\ \hline
                   0.1  &0.79$\pm$0.40&3.59/0.86	& 5.23$\pm$3.79                                                     & 20.97/4.13  \\ \hline
                   0.05 &0.81$\pm$0.39&4.00/0.87	&  5.29$\pm$3.52                                                    &   21.48/4.50  \\ \hline  
\end{tabular}
\end{table}
\subsection{Impact of the Binning Size of Training}
\label{evaluation_rss_threshold}
Next, we examined the influence of accuracy with respect to various binning sizes $r$. 
We chose RSS threshold $\vartheta_r=-70$ and similarity threshold $\vartheta_s=0.7$. 
To evaluate accuracy under impact of different binning sizes, 
we set $r$ to the following values $r=\{0.05, 0.1, 0.2, 0.4, 0.6, 0.8, 1.0\}$. 
In our approach, the uncertainty model is trained with all the collected data. 
We show a comparison of the results in Table \ref{table_binning}. 
This table shows that the best choice of $r$ is $0.2$. 
The covariance estimated with a large $r$ is usually too large 
to compensate for the error from the odometry.
Optimizing $r$ gives an improvement of 25.2\% (0.77m and 1.03m for $r=0.2$ and $r=1.0$ respectively) for Tango-based PDR 
and 16.6\% (5.19m and 6.23m for $r=0.2$ and $r=1.0$ respectively) for step counter-based PDR.
Covariance of the loop is a key to optimize the pose graph, 
as it is the only information to measure how close 
the two locations are in a loop, 
therefore, a careful examination of the parameter will lead to an improvement of the accuracy.
The covariance added here (see Figure \ref{fig:sim_distance}) is much smaller as compared to the 
accumulated odometry error 
(notice that the maximum positioning error of Tango and step counter in Table \ref{table_rss} are 31.55m and 41.23m respectively). 
This is why we are still able to correct the accumulated odometry error. 
The approach presented here provides a way to automatically 
calibrate the uncertainty model with the odometry measurement. 
The model generated in different environments might be slightly different, as shown in Figure \ref{fig:sim_distance}. 
To evaluate the accuracy under the impact of different similarity models, 
Table \ref{table:accuracy_diff_bld} compared the results using the similarity models produced from two different buildings, namely Building1 (the one used for the verification of the positioning accuracy) 
and Building2 regarding to Tango-based approach. 
As can be seen from this table, the two models provide similar positioning accuracy, 
which proves our assumption that the similarity model can be applied to different environments.

\begin{table}[]
\centering
\caption{
comparison of mean positioning accuracy under different similarity models trained at two different buildings regarding Tango-based pedestrian dead reckoning system.
}
\label{table:accuracy_diff_bld}
\begin{tabular}{|c|c|c|c|c|c|c|c|}
\hline
\multirow{2}{*}{Model} & \multicolumn{7}{c|}{Binning size $r$}            \\ \cline{2-8} 
                        & 0.05 & 0.1  & 0.2  & 0.4  & 0.6  & 0.8  & 1.0  \\ \hline
Bld1               & 0.81 & 0.79 & 0.77 & 0.81 & 0.84 & 0.85 & 1.03 \\ \hline
Bld2               & 0.84   & 0.81   & 0.78   & 0.85   & 0.86   & 0.91   & 1.07  \\ \hline
\end{tabular}
\end{table}

\subsection{Impact of Turning Detection and Matching}
\label{evaluation_motion_constraints}
\begin{figure}
  \centering     
        \subfigure[Positioning accuracy of Tango-based pedestrian dead reckoning system under different settings of $w$]{
\label{fig:trajectory}
        \includegraphics[width=0.48\textwidth]{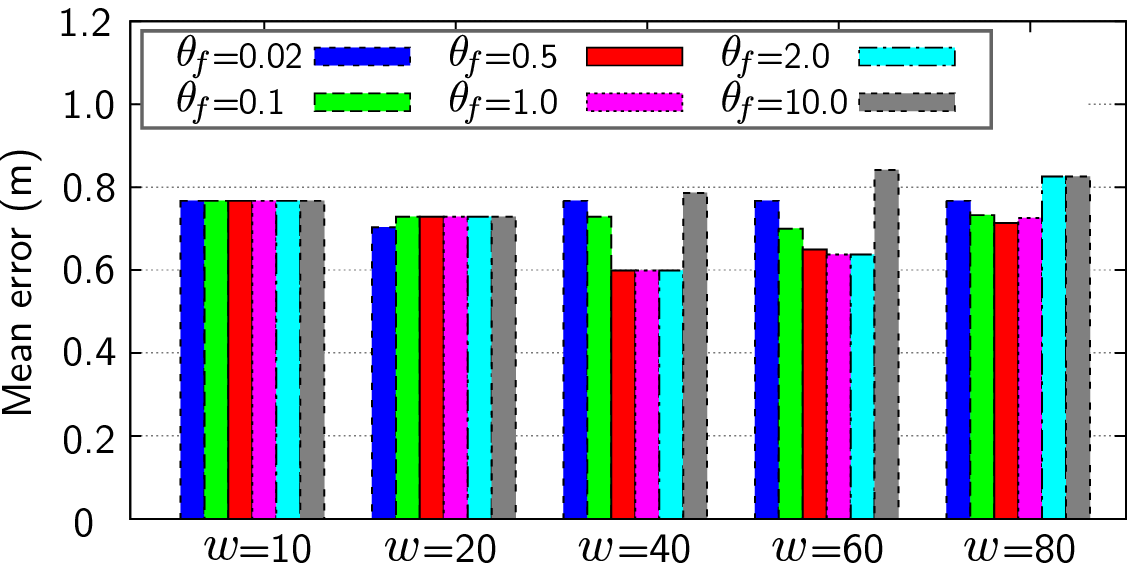}
        }
         \subfigure[Positioning accuracy of step counter-based pedestrian dead reckoning system under different settings of $w$]{
\label{fig:trajectory}
        \includegraphics[width=0.48\textwidth]{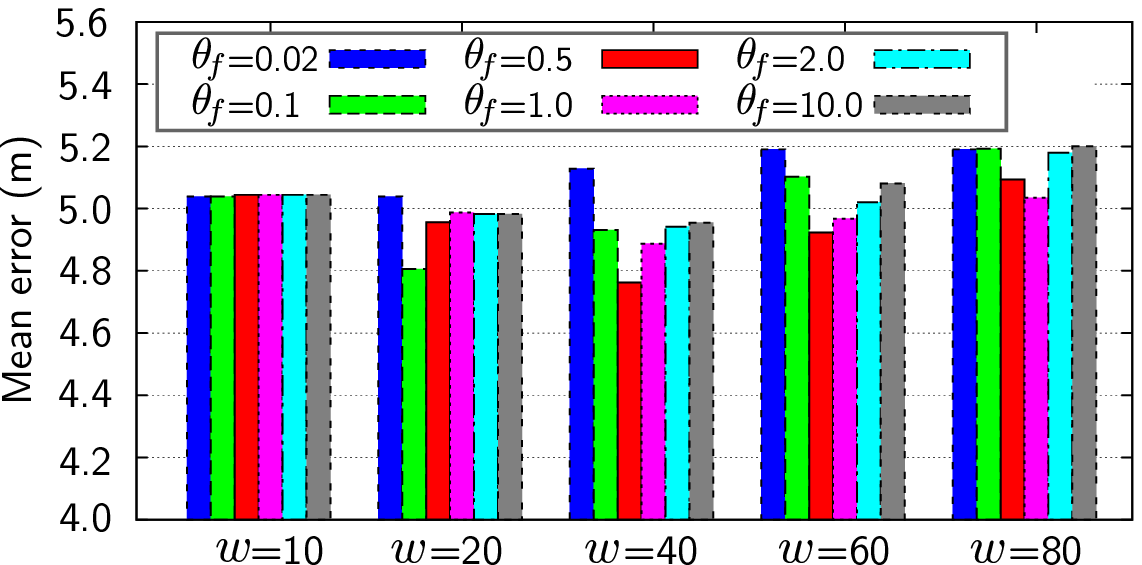}
        }    
   \caption[Turning detection]
{The impact of turning detection and matching on average positioning accuracy with two different pedestrian dead reckoning systems 
under various settings of window size $w$ and $\vartheta_f$.
}
\label{fig:accuracy_under_turning_detection}
\end{figure}

We compared the accuracy with and without the integration of turning features in the next series of experiments. 
We fixed RSS threshold $\vartheta_r=-70$, similarity threshold $\vartheta_s=0.7$, and binning size $r=0.2$. 
We varied the setting of $w$ and $\vartheta_f$ to evaluate their impact on the accuracy. 
Figure \ref{fig:accuracy_under_turning_detection} shows a comparison of the results.
This figure shows that 
the accuracy can be improved by additional integration of turning features:
we obtain an improvement of 22.1\% for the Tango-based pedestrian dead reckoning 
(from 0.77m to 0.6m with $w=40$ and $\vartheta_f=0.5$) 
and 8.3\% for step counter-based dead reckoning (from 5.19m to 4.76m with $w=40$ and $\vartheta_f=0.5$).
The improvement of Tango is slightly higher as compared to the step counter-based system.
Since the odometry error of Tango is smaller than step counter, 
the turning feature here exhibits great capability to correct the drift error of odometry. 
However, for step counter-based PDR, the error is dominated by the odometry 
and the turning feature shows less improvement to the accuracy as compared to Tango-based PDR. 

The number of turnings detected with different settings of $w$ is shown in Table \ref{table:turnings_detected}.
The constraints are mostly from the fingerprinting matching. 
As can be seen from Table \ref{table:turnings_detected}, only 26 turnings are detected with $w=40$, 
which is much less than the number of fingerprinting-based constraints (715 for Tango-based approach with $\vartheta_s=0.7$ as shown in Table \ref{table_sim}).
Due to the low sampling rate, 
Tango-based PDR leads to a small number of turnings as compared to the step counter-based PDR. 
A large value of $w$ leads to an increasing number of turnings detected.
From Figure \ref{fig:accuracy_under_turning_detection}, we can also observe that
a window size $w=40$ leads to the best accuracy for both systems. 
A too high or too small $w$ obviously results in a less improvement to the accuracy.

In addition, 
Figure \ref{fig:accuracy_under_turning_detection} shows that 
a too large or too small $\vartheta_f$ leads to a decrease of the accuracy. 
A suitable fitness threshold $\vartheta_f$ will help to remove the false turning-based loop closures. 
A small $\vartheta_f$ will not be able to identify true matched turnings 
and results in a small number of turning-based loops 
and therefore has less strength to improve the accuracy.
A large $\vartheta_f$ produces too many false turning-based loops, 
which leads to a deterioration of the result.
As an example, for step counter-based PDR with $w=40$, 
a setting of $\vartheta_f=0.5$ gives an accuracy of 4.76m, 
which produces an improvement of 3.8\% and 7.0\% when compared to $\vartheta_f=10.0$ (4.95m) and $\vartheta_f=0.02$ (5.12m), respectively.


\begin{table}[]
\centering
\caption{
comparison of number of turnings detected with two different pedestrian dead reckoning systems under various settings of window size $w$.
}
\label{table:turnings_detected}
\begin{tabular}{|c|c|c|c|c|c|}
\hline
Window size $w$           & \multicolumn{1}{c|}{10} & \multicolumn{1}{c|}{20} & \multicolumn{1}{c|}{40} & \multicolumn{1}{c|}{60} & \multicolumn{1}{c|}{80} \\ \hline
Turnings (Tango)        &         0               &        6                 &        26                 &    45                     &       78                  \\ \hline
Turnings (step counter) &                0        &              423           &             1447            &       2666                  &    3973                     \\ \hline
\end{tabular}
\end{table}


\subsection{Computational Time}
\label{evaluation_rss_threshold}
Finally, we evaluated the time consumption at each stage of our approach.
The results are listed in Table \ref{table:computational_time}. 
In our approach, we process the recordings in a batch matter, which means that we optimized the graph after all loop closures are identified.
An Intel Core i5-4200M CPU with 2.5GHz frequency and 4GB RAM is used to process the measurements. 
We set $\vartheta_r=-70$, $\vartheta_s=0.7$, $r=0.2$, $w=40$, and $\vartheta_f=0.5$. 
As can be seen from Table \ref{table:computational_time}, 
the entire data processing took 17.81 (6.08+5.73+5.93+0.07) seconds for tango-based dataset, which is almost 30 times faster than the data recording stage (544.75 seconds),
while the processing time is much longer for step counter-based dataset (approx. 129.39 seconds, i.e., 23.58+72.34+31.82+1.65), due to its higher sampling rate.
Additionally, 
optimization of the graph only took less than two seconds (0.07 and 1.65 seconds for Tango and step counter-based system respectively).
In our current offline implementation, 
we compute the similarity of all pairs of Wi-Fi measurements in the entire dataset to find the potential loop closures. 
This module consumes too much time (72.34 seconds for the step counter version as shown in Table \ref{table:computational_time}), 
and thus cannot be used for real time pose estimation.
However, 
we believe the computation can be further optimized to make online implementation possible, 
e.g. we only need to compare the current Wi-Fi measurement with the previous measurements for the loop closure detection 
(estimated to be 72.34/2179 = 0.03 seconds), 
which is less than the current Wi-Fi sampling rate of one second interval. 
Nonetheless, we believe further optimization is needed to ensure online implementation in real time.

When more users are involved in the experiment, it will take longer time to run the algorithm due to the increasing number of nodes and the constraints in the graph. 
Still, our approach is efficient when compared to vision-based approaches, 
as vision-based approach requires heavy computational resources due to feature extraction and feature matching. 
In addition, there can be data association problems which will result in the loop closure failure. On other hand, the MAC address of the AP is unique. 
It is not necessary to run the optimization algorithm for each new Wi-Fi measurement. 
We suggest to run the optimization when a loop closure is detected or a certain amount of loop closures has been identified. 
Solutions to reduce the computational time can be found in \cite{life_long_SLAM} \cite{Long_term_mapping}.
One might notice that the model training described in Section\,\ref{sect_model_training} takes long time. 
But this phase can be performed offline, and the learned model can be saved and applied to other users or different environments once it is generated.

\begin{table}[]
\centering
\caption{
Evaluation of the computational time (in seconds) in each stage using two different pedestrian dead reckoning systems.
}
\label{table:computational_time}
\begin{tabular}{|c|c|c|}
\hline
\multicolumn{1}{|c|}{\multirow{2}{*}{\begin{tabular}[c]{@{}c@{}} Stage\\  \end{tabular}}} & \multicolumn{2}{c|}{duration(s)}                               \\ \cline{2-3} 
\multicolumn{1}{|c|}{}                                                                      & \multicolumn{1}{c|}{Tango} & \multicolumn{1}{c|}{Step counter} \\ \hline
Data recording (time per track) & 544.75 & 544.75        \\ \hline
\begin{tabular}[c]{@{}c@{}}Model training+\\ variance computation \end{tabular} & 6.08 & 23.58 \\ \hline
Loop closure detection  &   5.73  & 72.34 \\ \hline
Turning detection and matching &  5.93   & 31.82 \\ \hline
Pose graph optimization &  0.07   & 1.65 \\ \hline
\end{tabular}
\end{table}

\section{Conclusions and Future Work}
\label{conclusions}
In this paper, we presented a novel approach for collaborative simultaneous localization and radio fingerprint mapping (C-SLAM-RF) 
in unknown environments. 
The proposed system makes use of a pedestrian dead reckoning system 
and the RSS measurement from surrounding wireless access points. 
We further incorporate the motion features to improve the accuracy of the system. 
The proposed approach does not require any knowledge of the map and locations of the access points.
The performance of our approach is evaluated in a large scale environment 
under two pedestrian dead reckoning systems with different motion tracking accuracies. 
Our results reveal that the accuracy of a pedestrian tracking system plays an important role in the accuracy of our approach. 
We obtained an accuracy of 0.6m and 4.76m for Tango and step counter-based pedestrian dead reckoning systems, respectively.
The quality of the radio map will increase with more users involved in collecting the measurements due to the crowdsourcing nature of the proposed approach. 
One of our future work is to enhance the accuracy of step counter-based PDR 
by stride length estimation and the fusion of gyroscope measurement.
Another direction would be the evaluation of the indoor positioning accuracy by applying the radio map constructed from our SLAM system. 
\bibliographystyle{IEEEtran}
\bibliography{literatur}

\end{document}